\documentclass[usenatbib,useAMS]{mn2e}
\usepackage{epsfig,amsmath,times}
\def\gsim{\;\rlap{\lower 2.5pt
 \hbox{$\sim$}}\raise 1.5pt\hbox{$>$}\;}
\def\lsim{\;\rlap{\lower 2.5pt
   \hbox{$\sim$}}\raise 1.5pt\hbox{$<$}\;} 

\title{{\it Chandra} Observations of the Nucleus of M33}
\date{March 2002}
\pubyear{2002} \volume{000} \pagerange{0-0}

\author[G. Dubus and R. E. Rutledge]
{Guillaume Dubus and Robert E. Rutledge \\ California Institute of Technology, MS 130-33, Pasadena, CA 91125, U.S.A.}

\begin{document}
\maketitle

\label{firstpage}

\begin{abstract}
The nearby galaxy M33 hosts the most luminous steady X-ray source in
the Local Group. The high spatial resolution of {\it Chandra} allows
us to confirm that this ultra-luminous X-ray source is within the
nucleus and rule out at the 4.6$\sigma$ level a previously proposed
possible counterpart located 1\arcsec\ away. The X-ray spectrum is
well fitted by a disc blackbody with $kT_{\rm in}$=1.18$\pm$0.02~keV
and $R_{\rm in}(\cos\theta)^{0.5}$=57$\pm$3~km, consistent with
earlier results from {\it ASCA} and {\it BeppoSAX}. The source flux is
steady between 1 and $10^4$~s ($<$3\% rms variability, 0.5-10 keV)
with a 0.5-10~keV luminosity of $1.5\times10^{39}$~erg~s$^{-1}$.  The
X-ray properties and an association with a radio source are
reminiscent of the galactic microquasar GRS~1915+105.
\end{abstract}


\section{Introduction} 

The nearby galaxy M33 ($d\approx 795$~kpc; \citealt{vdb}) hosts the
most luminous steady X-ray source in the Local Group. M33 X-8
dominates the X-ray emission from the galaxy with a 1-10~keV
luminosity of about 1.2$\cdot 10^{39}$~erg~s$^{-1}$, ten times more
than that of the other X-ray sources in the field. The ultra-luminous
X-ray source (ULX) has been detected at this level since its discovery
by {\it Einstein}.
\citep{long,markert,gottwald,peres,takano,lc96,parmar}.  Observations
obtained with the {\it ROSAT} High Resolution Imager (HRI) showed that
X-8 is point-like and coincident with the optical nucleus of M33 at
the HRI 5\arcsec{} resolution \citep{schulman}. These also revealed a
106 day $\sim$20\% modulation of the X-ray flux from M33 X-8
\citep{106}.

The nature of the X-ray emitting object is rather puzzling. The
measured velocity dispersion of the nucleus places an upper
limit of 1500~M$_\odot$ on the mass of a putative massive central
black hole \citep{gebhardt}.  A low luminosity active galactic nucleus
(AGN) is therefore unlikely. Indeed, the UV/optical spectrum of the
nucleus shows no signs of AGN-type activity and is best modelled by a
combination of two starbursts at 40 Myrs and 1 Gyrs (\citealt{lcd} and
references therein).

The 106 day modulation suggests a single object is responsible for
most of the X-ray emission, perhaps a persistent X-ray binary with a
10 M$_\odot$ black hole radiating at the Eddington luminosity. {\em
Hubble Space Telescope} observations place a limit of 18\% on the
contribution of a point source to the total far-UV flux from the
nucleus \citep{dubus}.  The implied ratio $L_\rmn{opt}$/$L_\rmn{X}\sim
0.05$ would be in line with expectations from a low-mass X-ray
binary. The counterpart would then be mostly lost in the glare of the
nucleus. However, there remains a distinct possibility that M33 X-8 is
associated with a UV bright star located 1\arcsec{} to the NNW
\citep{dubus}. The star has a O9III spectrum with no outstanding
feature linking it to the powerful X-ray source but cannot be ruled
out at the {\it ROSAT} HRI resolution \citep{lcd}.

Most studies have assumed M33 X-8 is associated with the nucleus and
used this to register the X-ray images. {\it Chandra} offers the
opportunity to locate M33 X-8 to unprecedented accuracy and settle
whether this source is within the nucleus or not. Three observations
of M33 are available in the {\it Chandra} archive (Table~1). We used
the first two to accurately position M33 X-8 with respect to {\it HST}
Planetary Camera (PC) images of the nucleus (\S2). The third, non
piled-up observation is used to study the X-ray spectrum and timing
behaviour of M33 X-8 (\S3 and \S4). Our results are discussed in the
last section.

\begin{table}
\caption{{\em Chandra} Observations of M33}
\begin{tabular}{@{}lllllll@{}}
Det. & ObsId & Seq.\# & Date & Exp. & offset\\
\\
ACIS-I & 1730 & 600145 & 7/12/2000 & 49.4~ks &  0\arcmin   \\
ACIS-S & 786  & 600089 & 8/30/2000 & 46.2~ks &  0\arcmin \\
ACIS-S & 787  & 600090 & 1/11/2000 & 9.2~ks  &  7.7\arcmin\\
\end{tabular}
\medskip

Det. is the detector used; exp. is the total integration time; offset is the distance of M33 X-8 from the telescope axis. The CCD frame time was 3.2~s for the first two observations and 1.0~s for the last, offset observation.
\end{table}

\section{Astrometry}

The 1-2\arcsec\ precision of the {\it Chandra} pipeline aspect
solution\footnote{http://cxc.harvard.edu/cal/ASPECT/celmon/index.html}
is insufficient for our purposes. To improve on this, we identified
X-ray sources within the field-of-view and cross-correlated these to
optical, IR and radio sources with accurately known positions.

\subsection{X-ray position of the nucleus} 

M33 X-8 is severely affected by pile-up in the two 50~ks ACIS
observations.  The count rate in the ACIS-S observation is 0.5
counts/frame compared to 3.0 counts/frame in the 10~ks non piled-up
10~ks ACIS-S observation (see \S3 below).  The pile-up fraction is
therefore about 84\% which makes it impossible to study the radial
extension of the source to higher accuracy than the {\em ROSAT} HRI.
Observations with the {\em Chandra} HRC would be able resolve any
contribution from extended emission or faint point sources within
5\arcsec. The inner pixels of the ACIS-S (Fig.~1) are very piled-up
hence have a dearth of counts so that the 0.5-2~keV image is doughnut
shaped (standard G02346 events). Going to higher energies reduces the effect
and there is a very well defined peak in both ACIS-S and I. 

CELLDETECT (CIAO v2.2.0.1\footnote{http://cxc.harvard.edu/ciao/})
locates M33 X-8 at (4098.1,4057.2)$\pm$0.1~pixels in the ACIS-I and at
(4063.0,4026.3)$\pm$0.5~pixels in the ACIS-S. To check these values,
we redefined the center as the location which minimizes the distances
of all the events within a 300 pixel wide box roughly centered on M33
X-8 (this assumes circular symetry and avoids the problem posed by a
pile-up doughnut). The values obtained by selecting only the events
with energies above 1, 2 or 5~keV are all consistent with the
CELLDETECT locations and errors which we therefore adopt.
 
\subsection{Optical position of the nucleus} 
The optical position of the nucleus we adopt is the 2MASS position
which is tied to the Tycho reference frame. The (compact) nucleus is
detected with $J=12.08\pm0.04$ in this survey.  At
$\alpha~(\rmn{J2000})~1^{\rm{h}}{33}^{\rm{m}}50\fs89$ and
$\delta~30^\circ39\arcmin36\farcs75$ with an uncertainty of 0\farcs15
this is, as far as we know, the best available absolute position.  The
position of the close O9III star was then found relative to the
nucleus using the UV {\it HST} PC image described in \cite{dubus}. We
find $\alpha~1^{\rm{h}}{33}^{\rm{m}}50\fs85$ and
$\delta~30^\circ39\arcmin37\farcs65$ to the 2MASS uncertainty.

\begin{figure}
\centerline{\epsfig{file=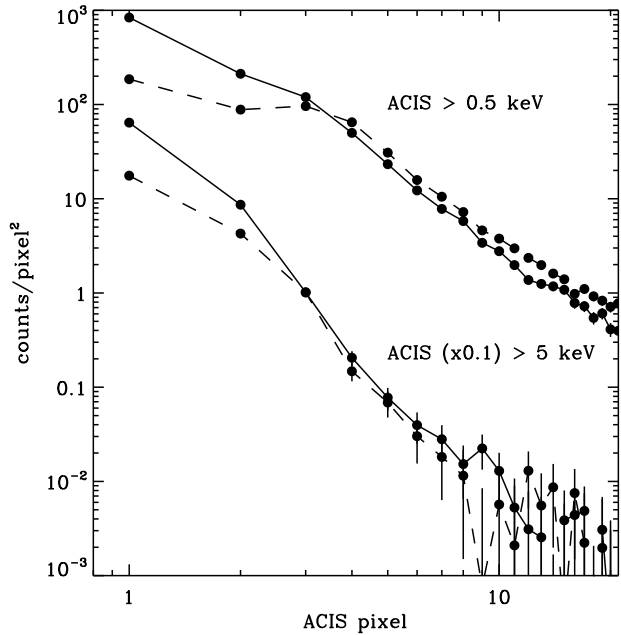}}
\caption{Radial profiles of M33 X-8 in the two piled-up 50~ks ACIS-I
(solid lines) and S (dashed line) observations
(1~pixel=0.492\arcsec). All G02346 events with energies $>0.5$~keV are
used in the top two profiles (9900 and 6600 counts respectively for
ACIS-I and S) while only those above 5~keV are used in the bottom
profiles (3100 and 1200 counts; profiles are shifted down by a factor
10 for clarity). The profiles are (background subtracted)
area-averaged number of events within x and x-1 pixels of the adopted
centers (\S2.1). Pile-up appears most clearly as a flattening of the
ACIS-S profile at low energies.}
\end{figure}

As an independent check, we cross-correlated the brightest stars in
the various {\it HST} PC images of the nucleus \citep{dubus} with the
M33 star catalogue produced by the DIRECT project
\citep{macri_cat}. The DIRECT positions are tied to the USNO-A2
reference frame with an accuracy of 0\farcs4-0\farcs7. We found 4-6
stars (depending upon the sky orientation of the PC chip) that
correlated well with DIRECT positions, all lying at the same offset
within 0\farcs1-0\farcs4 of each other. We found a mean corrected
position for the nucleus of $\alpha~1^{\rm{h}}{33}^{\rm{m}}50\fs90$
and $\delta~30^\circ39\arcmin37\farcs14$. This is consistent with the
2MASS position within the limits of the DIRECT astrometric solution.

\subsection{X-ray source list and cross-correlation}
\begin{figure*}
\centerline{\epsfig{file=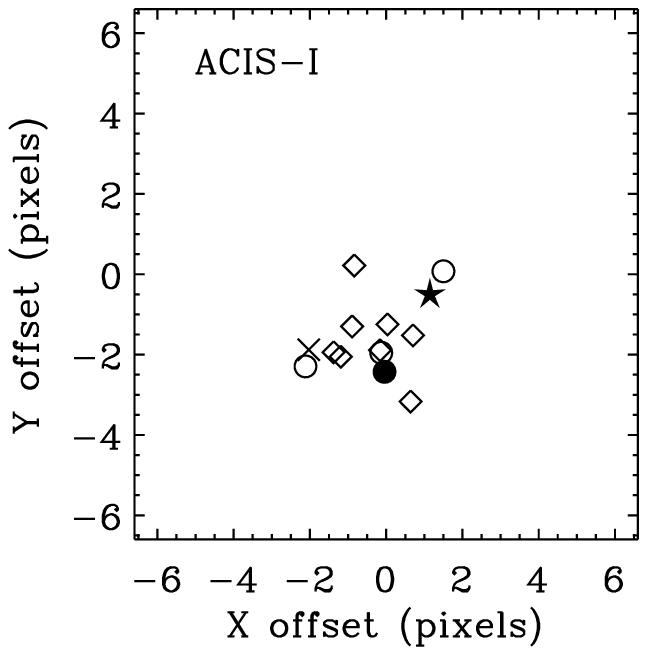}\qquad\epsfig{file=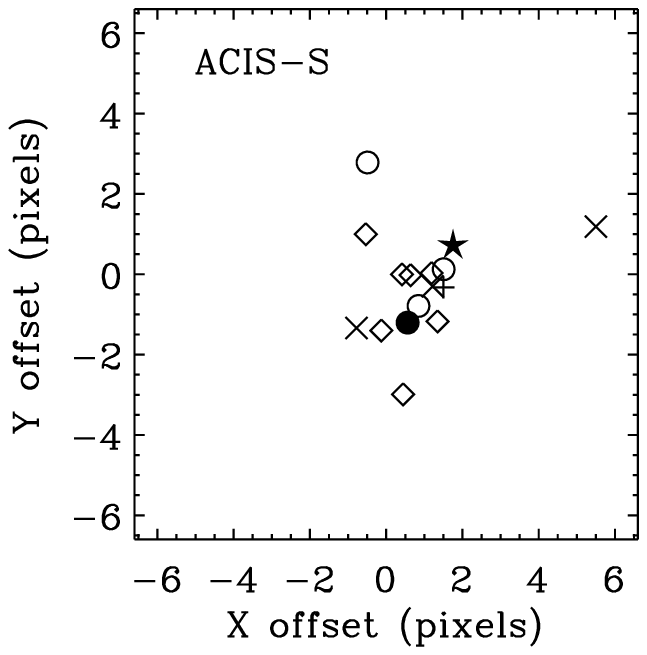}}
\caption{Plot of the offsets in ACIS pixels (1~pixel=0.492\arcsec)
between X-ray sources and their matches in the USNO ($\times$), Tycho
($+$), 2MASS ($\circ$), and radio ($\diamond$, \citealt{gordon99})
catalogues (see Table 1). The offset between the {\em Chandra}
position of X-8 and the 2MASS position of the nucleus is indicated by
a $\bullet$. The offset between X-8 and the UV bright star $\sim
1$\arcsec\ NNW of the nucleus is shown by a $\star$. The radio
counterparts cluster at an offset consistent with the ULX being associated
with the nucleus ($\bullet$) rather than with the nearby star
($\star$).}
\end{figure*}
\begin{table}
\caption{All possible matches within 3\arcsec{} of X-ray sources (Fig.~2).}
\begin{tabular}{@{}llllrrl}
Source & Cat. & $\alpha$ & $\delta$ & $\Delta$X & $\Delta$Y & err\\
 &  & \multicolumn{2}{l}{(degrees, J2000)} & \multicolumn{2}{c}{(pixels)}& (\arcsec)\\
\\
\multicolumn{7}{l}{ACIS-I observation (ObsId 1730)}\\
\\
HP102 &  2MASS & 23.46205 & 30.66021 &-0.0& -2.4 & 0.16\\%& GC\\
      & GDK102 & 23.46204 & 30.66037 & 0.0& -1.2 & 0.60 \\%& GC\\
 HP84 &  2MASS & 23.42445 & 30.64703 &-0.1& -2.0 & 0.16\\%& \\
 HP74 &  GDK64 & 23.39958 & 30.60785 & 0.7& -1.5 & 0.26 \\%& SNR\\
HP130 & GDK148 & 23.54462 & 30.70671 &-1.2& -2.1 & 0.40\\ %& SNR\\
 HP62 &   USNO & 23.37106 & 30.70483 &-2.0& -1.9 & 0.19\\%& \\
      &  GDK50 & 23.37096 & 30.70482 &-1.4& -1.9 & 0.42\\ %& SNR\\
CXO1  &  2MASS & 23.34009 & 30.65008 & 1.5 &  0.1 & 0.46\\%& \\
      &  2MASS & 23.34066 & 30.64976 &-2.1 & -2.3 & 0.46\\%& \\
%     &  GDK38 & 23.34046 & 30.65010 &-0.8 &  0.2 & 1.26\\%& B-STR\\
      &  GDK38 & 23.34046 & 30.65010 &-0.8&  0.2 & 1.26\\%& \\
HP106 & GDK112 & 23.47858 & 30.55301 &-0.2& -1.9 & 0.26\\%& SNR\\
%HP80 &  GDK67 & 23.40621 & 30.78870 & 0.6& -3.2 & 0.52\\%& $<$AGN$>$ B-STR\\
 HP80 &  GDK67 & 23.40621 & 30.78870 & 0.6& -3.2 & 0.52\\%& AGN\\
 HP67 &  GDK57 & 23.38008 & 30.55939 &-0.9& -1.3 & 0.27\\%& SNR\\
\\
\multicolumn{7}{l}{ACIS-S observation (ObsId 786)}\\
\\
HP102 &  2MASS & 23.46205 & 30.66021 & 0.6 & -1.2 &  0.29 \\%& GC\\
      & GDK102 & 23.46204 & 30.66037 & 0.6& -0.0 & 0.65\\%& GC\\
% HP98 & GDK100 & 23.46025 & 30.63937 & 1.3& -1.2 & 0.25%& $<$agn$>$ B-NOE\\
 HP98 & GDK100 & 23.46025 & 30.63937 & 1.3& -1.2 & 0.25\\%& AGN\\
 HP84 &  2MASS & 23.42445 & 30.64703 & 0.8& -0.8 & 0.15 \\%& SSS\\
 HP62 &  GDK50 & 23.37096 & 30.70482 &-0.1& -1.4 & 0.42\\%& SNR\\
      &   USNO & 23.37106 & 30.70483 &-0.8& -1.3 & 0.20\\%& \\
HP106 & GDK112 & 23.47858 & 30.55301 & 0.4& -0.0 & 0.22\\%& SNR\\
 HP87 &  Tycho & 23.43076 & 30.77521 & 1.5& -0.3 & 0.36\\%& \\
      &  2MASS & 23.43076 & 30.77527 & 1.5&  0.1 & 0.38\\%& \\
      &   USNO & 23.43081 & 30.77521 & 1.2& -0.3 & 0.38\\%& \\
      &   USNO & 23.43013 & 30.77541 & 5.5&  1.2 & 0.38\\%& \\
%CXO-2 &   GK51 & 23.37192 & 30.76057 & 0.4 & -3.0 & 0.41 %& B-NOE\\
CXO2 &   GDK51 & 23.37192 & 30.76057 & 0.4& -3.0 & 0.41\\%& \\
% HP80&  GDK67 & 23.40621 & 30.78870 & 1.2&  0.0 & 0.48 %& $<$AGN$>$ B-STR\\
 HP80 &  GDK67 & 23.40621 & 30.78870 & 1.2&  0.0 & 0.48\\%& AGN\\
CXO3 &  2MASS & 23.49518 & 30.83467 &-0.5&  2.8 & 0.51\\%& \\
 HP64 &  GDK52 & 23.37279 & 30.81954 &-0.5&  1.0 & 0.94\\%& SNR\\
\end{tabular}

\medskip
HP refers to the \cite{haberl} catalog of ROSAT sources and is used to
identify the sources (column 1). CXO refers to {\it Chandra} sources
which were not previously detected. GDK refers to the \cite{gordon99}
catalog of radio sources in M33 (their Table~1). For each match we
list the catalogue (column 2) and position in this catalogue (columns
3 \& 4). $\Delta$X and $\Delta$Y (columns 5 \& 6) are the offsets in
ACIS pixels between the X-ray and matched source position (plotted in
Fig.~2). Column 5 is the offset error $(\sigma^2_{\rm X}+\sigma^2_{\rm
cat})^{1/2}$, where $\sigma^2_{\rm X}$ is the error in the X-ray
source position (given by CELLDETECT, see \S2.3) and $\sigma^2_{\rm
cat}$ is the error in the position of the correlated source (given in
the relevant catalogue, see \S2.3).
\end{table}
We applied CELLDETECT to identify X-ray sources in the two 50~ks
observations. The detection treshold was 4$\sigma$, resulting
in 51 (77) possible sources in the ACIS-I (ACIS-S) dataset. We
compared this list with the {\it ROSAT} X-ray catalogue of
\cite{haberl} to identify previously known X-ray sources. {\it ROSAT}
and {\it Chandra} sources lying within 5\arcsec{} of each other were
considered identical and are thereafter identified by their
number in Table~1 of \cite{haberl}.

We cross-correlated the X-ray source list with the Tycho-2
\citep{tycho}, USNO-A2 \citep{usno} and 2MASS \citep{2mass}
catalogues. We also used a subset \citep{macri_var, directx7} of the
DIRECT catalogue containing variable optical sources (excluding
cepheids) and the radio source catalogue of \cite{gordon99}. The
latter contains optically confirmed supernova remnants (SNR) in M33
and stellar-like background radio sources. The radio positions are
accurate with respect to the ICRS frame to 7\arcsec/2$\sigma$, where
7\arcsec\ is the beam size and $\sigma$ is the signal-to-noise ratio
of the source.

We started by searching for all possible counterparts in the
catalogues within 10\arcsec{} of each X-ray source and plotting their
pixel offsets relative to the {\it Chandra} position. A given X-ray
source can be associated with several sources from the same or
different catalogues. In the plot, true counterparts will cluster at
the offset of the X-ray astrometry (the orientation error and ACIS
chip distortion are negligible), while spurious identifications will
appear uniformly distributed (the field of view is much greater than
the searched correlation radius). The spread of the cluster is a
measure of the accuracy of the astrometric solution.  There was an
obvious concentration of matches around the offset between X-8 and the
2MASS position of the nucleus (Table~2 and Fig.~2).  Note that some of
the matches may refer to the same object such as the USNO and Tycho
match for HP87. None of the DIRECT variable sources matched an X-ray
source within 3\arcsec{} although we note that a star with a 3.4 days
optical period (D33J013334.8+303211.4, \citealt{directx7}) lies within
10\arcsec{} of the 3.45 day eclipsing X-ray source X-7
\citep{x7}.

\subsection{Position of the nuclear X-ray source}
The clustering of matches around the offset between the 2MASS position
of the nucleus and the {\it Chandra} position of X-8 is particularly
clear for the radio correlations (Fig.~2). For uniformly distributed
sources, the probability $p$ of having an X-ray source within $d$
arcsec of a possible counterpart is $p \approx (d/D)^2$ if $d\ll D$
and where $D$ is the radius of the field of regard. Correlating $N$
X-ray sources with $M$ possible counterparts in the field of view, the
probability $P$ of observing $C$ or more sources within $d$ arcsec of
a counterpart is $$P=1- \sum_{i=0}^{C-1} \binom{MN}{i} p^i
(1-p)^{MN-i}$$ For $C=8$ matches with the radio catalogue,
$d=3$\arcsec{}, $N=32$, $M=126$ and $D\approx8$\arcmin, the
probability of seeing a cluster of 8 associations within 3\arcsec{} is
$\lsim 10^{-11}$. The same calculation for the ACIS-S observation
($C=7$, $N=57$, $M=117$) yields a probability $\lsim 10^{-8}$. We
conclude the radio associations are very likely to be true
counterparts (in contrast with the USNO, 2MASS and Tycho associations
for which the same calculation does not rule out chance
superpositions).

We now estimate the probability that the offsets given by the radio
counterparts are consistent with the offset between the 2MASS position
of the nucleus and X-8 ($\bullet$ in Fig.~2) or the nearby star
position and X-8, plotted as a $\star$ in Fig.~2). A $\chi^2$ test on
the measured distances between the radio and nucleus offsets yields a
probability 0.28 (0.53) that these are consistent for the ACIS-I
(ACIS-S) data. The probability for the radio offsets to be consistent
with the nearby star are $2\cdot 10^{-6}$ (ACIS-I) and $4\cdot
10^{-4}$ (ACIS-S). The ACIS-S result is less conclusive than the
ACIS-I because of the higher uncertainty in the position of X-8 in the
ACIS-S ($\pm$0.5 pixel compared to $\pm$0.1 pixel, see \S2.1). We
conclude the nearby star is ruled out as a possible counterpart at the
4.6$\sigma$ level and that M33 X-8 is at the 2MASS position of the
nucleus with an uncertainty of 0\farcs6 (the mean weighted distance of
the radio offsets for both ACIS-I and S).

There is a radio source associated with the nucleus. This source
(GDK102) has $F_\rmn{20cm}=0.6\pm0.1$~mJy, $F_\rmn{6cm}=0.2\pm0.1$~mJy
is located 0\farcs6 away from the 2MASS position (and 0\farcs4 away
from the mean radio offset) and is a possible counterpart to M33
X-8. Using the results of the above analysis, we note the association
of X-ray source HP84 with 2MASS~0133418+303849 ($J=12.62\pm0.02$) is
likely to be real. HP84 is classified as a supersoft X-ray source by
\cite{haberl}.

\section{X-ray Spectrum}
We performed spectral analysis using data from the ACIS-S observation
ObsId 787.  We extracted 27259 counts from a circle within 20\arcsec\
of the source position for a total of 3.06 counts/frame.  The source
was 7.7\arcmin\ off-axis, at which the the 50\% enclosed energy
function (EEF) at 1.5 keV is 4\arcsec\ vs. 0.3\arcsec on-axis,
resulting in the source covering an area larger by a factor $\sim$170,
to 0.02 counts/frame/PSF, at which the estimated pileup fraction
should be negligible ($<$1\%).  We confirmed this by comparing the
distribution of photon event grades (for grades 0, 2, 3, 4 and 6)
below 2 keV and in the 2-8 keV range with those of a non-piled up
X-ray observation (of ZW3146, ObsId 1651, Seq. Num 800119) and found
that they were roughly consistent, though not identical; however, both
were similar to the branching ratio observed from a non piled-up
observation of PG1634-706
\footnote{http://asc.harvard.edu/cal/Hrma/hrma/misc/oac/psf2/index.html},
and so pileup does not appear to be significant during this
observation.  Background counts were taken from an annulus of inner-
and outer- radii of 25\arcsec\ and 100\arcsec.  The expected number of
background counts in the source region is 167$\pm$3.  The background
subtracted source countrate is then 2.92+/-0.02 c/s.

Using XSPEC v11.0.1 \citep{xspec}, we produced a pulse-height-analyser
(PHA) spectrum of the extracted counts data between 0.6-9 keV, binning
at 130 eV between 0.6 and 2.0 keV, at 260 eV between 2.0 and 5.4 keV,
and with two energy bins between 5.4 and 9.0 keV.  We included a 5\%
systematic uncertainty in the fits.  Best-fit parameters for
acceptable models are in Table\ref{tab:chandra}.

\begin{figure}
\centerline{\epsfig{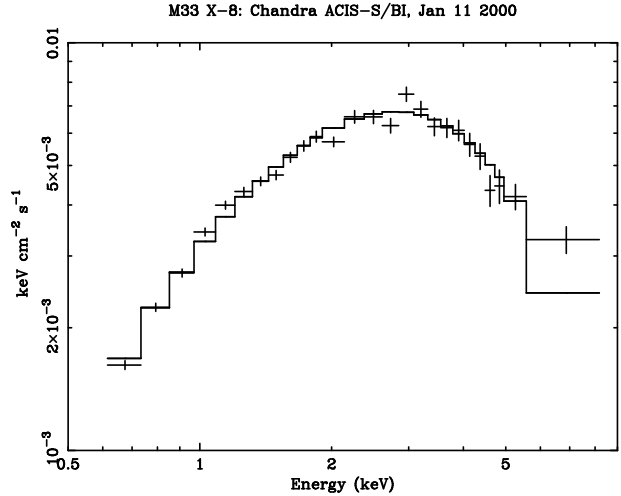}}
\caption{Best-fit absorbed disc-blackbody model (solid-line histogram), with
the observed {\em Chandra}/ACIS-S spectrum of M33 X-8 (crosses),
corrected for the detector response, in $\nu F_\nu$.  The deviation of
the data points near 3 keV from the model are not significant
(2$\sigma$); the excess near 6-9 keV deviates from the best-fit model
at the 3$\sigma$ level, and is consistent with possible contributions
from pile-up at the $\sim$1\% level, which are not accounted for in
the best-fit spectrum.}
\end{figure}

An absorbed powerlaw spectrum ({\tt wabs*powerlaw}) is rejected, with
a value of reduced chi-squared statistic $\chi^2_\nu$=4.12 for 23
degrees of freedom (dof), with a null-hypothesis probability of
$10^{-10}$.  The power-law spectrum over-predicts counts in the 2-4
keV range, and under-predicts them in the 4-9 keV range.  An absorbed
blackbody spectrum is also statistically rejected ($\chi^2_\nu=21.4$
for 23 dof).

We find a moderately acceptable fit ($\chi^2_{\nu}$=1.35) for an
absorbed thermal bremsstrahlung spectrum. An absorbed disc blackbody
spectrum produces a better fit to the data ($\chi^2_{\nu}$=0.95), with
an X-ray column density $N_H=(6\pm1)\times10^{20} {\rm cm}^{-2}$,
consistent with galactic absorption only
\footnote{W3NH at http://heasarc.gsfc.nasa.gov/ \citep{dickey90}}.
The disc blackbody inner temperature $kT_{\rm in}$ and inner-radius
$R_{\rm in}\sqrt{\cos{\theta}}$ are identical to the values found in
earlier observations with {\it ASCA} \citep{takano} and {\it BeppoSAX}
\citep{parmar}. The flux (corrected for absorption) of the best-fit
spectrum is 2.0$\times10^{-11}\; {\rm erg}\;{\rm cm}^{-2}\; {\rm
s}^{-1}$, for a luminosity 1.5$\times10^{39}\; {\rm erg}\; {\rm
s}^{-1}$ (0.5-10 keV), consistent (including spectral uncertainty)
with previously measured values.  Unlike previous work, a power-law
component is not required by the data.  When we include a power-law
component in addition to the absorbed disc-blackbody model, holding
$\alpha=2$ fixed, the 90\% confidence upper-limit for the flux of this
component is 1.1$\times10^{-11} {\rm erg}\;{\rm cm}^{-2}\; {\rm
s}^{-1}$.

\begin{table}
\caption{\label{tab:chandra} {\em Chandra} Observation Spectral Parameters (0.6-9 keV)}
\begin{tabular}{@{}ll}
Thermal Bremsstrahlung\\
$N_H$ ($10^{22} {\rm cm}^{-2}$)	& 0.19$\pm$0.01   \\
$kT$			&	3.5$\pm$0.1  keV	\\
$\int n_e n_I dV$	&	(1.74$\pm$0.05) $\times 10^{62}\; {\rm cm}^{-3}$\\
$\chi^2_{\nu}$ (dof)		& 1.35 (23)		\\
& \\
Disc Blackbody\\
$N_H$ ($10^{22} {\rm cm}^{-2}$)	& 0.06$\pm$0.01   \\
$T_{\rm in}$	&	1.18$\pm$0.02 keV	\\
$R_{\rm in}\sqrt{\cos{\theta}}$	&	57$\pm$3 km (D/795 kpc)\\
$\chi^2_{\nu}$ (dof)		& 0.95 (23)		\\
& \\
\end{tabular}
\medskip

Assumed source distance $d=795$ kpc.  Uncertainties are
1$\sigma$.
\end{table}

\section{X-ray Variability}

We examined the data for long-term noise-like time variability.  We
binned the 27259 counts detected in the full {\em Chandra} pass band,
with 1.04104 sec time resolution, and took a Fourier transform of the
data. With the resulting Fourier components, we obtained
a power-density spectrum (PDS; see \citealt{rogg} for the use of PDS
for broad-band timing analysis) which we rebinned logarithmically.
Examination of the resulting PDS shows no evidence of power above that
expected from Poisson noise.  Modelling the power as a power-law of
slope=1 plus a constant (held fixed at the Poisson level) in frequency
($P(\nu)=A\, \nu^{-1} + C$), we find an upper-limit to the
root-mean-square variability of $<$3\% (90\% confidence; 0.0001-0.5
Hz).

When we binned the data by a factor of 1, 10 and 100, (in which the
average counts per bin are 2.938, 29.4, and 294 counts), we find that
the maximum counts per bin was 11, 48 and 330 counts, which is
consistent with the maximum number of counts expected from Poisson
fluctuations.  Thus, we find no evidence for intensity variability on
timescales between 1 and 10,000 sec.

\section{Discussion}

The high luminosity of M33 X-8 is not sufficient to rule out a neutron
star X-ray binary since magnetospheric accretion can result in
super-Eddington luminosities. This was observed e.g. in the high mass
X-ray binaries (HMXBs) SMC X-1 or A0535-66. A HMXB would have been an
attractive possibility should M33 X-8 have been associated with the
nearby 09III star. The {\it Chandra} localisation of X-8 rules this
out (\S2). X-8 could still be a HMXB hiding in the nucleus but this
would be surprising since the compactness and spectra of the nucleus
make it unlikely that it hosts $>10 M_\odot$ stars \citep{lcd}; a HMXB
evolves on the uncomfortably short thermal timescale of the donor star
($<10^5$~yrs for $M_2 >10 M_\odot$); the X-ray spectrum of X-8 is
unlike that of typical HMXBs which have flat power laws in the
0.5-10~keV range; and the power spectrum shows no evidence for
pulsations.

On the other hand, the X-ray spectrum and absence of short-term X-ray
variability below 1~Hz are highly reminiscent of the high or very high
states of galactic black hole candidates \citep{tl}. In both states
the spectrum is dominated by a disc blackbody with a weak contribution
from a steep power law. The very high state shows 3-10~Hz QPOs with a
few \% rms superposed on 1-10\% rms flat top noise in the 1-10~Hz
range. There is little variability in the high state. GRS~1915+105, a
microquasar in our Galaxy with a $\sim 14$~M$_\odot$ black hole
accreting from a $\sim 1.2$~M$_\odot$ K giant star \citep{greiner},
also has such soft, low variability states \citep{belloni}. The
optical emission from such a system would be lost in the nucleus of
M33.  Since its discovery in 1992, GRS~1915+105 has been steadily
detected with X-ray luminosities around or in excess of
$10^{39}$~erg~s$^{-1}$.

The analogy with GRS~1915+105 is even more attractive considering the
0.6~mJy (respectively 0.2~mJy) radio source detected by the VLA at
20~cm \citep[6~cm;][]{gordon99} which we show to be consistent with
the ultra-luminous X-ray source (\S2.4). GRS~1915+105 has been known
to show steady radio emission around 30-50~mJy at 2~cm, framed by
powerful ejection episodes where the emission from the relativistic
jets becomes $\ga$ 500~mJy \citep{fender}. The radio source in M33 is
more powerful than that; but a small change in the beaming angle of
the jet ($\approx 70^\circ$ from the line-of-sight for GRS~1915+105)
could easily increase the observed radio flux. Follow-up observations
of the radio source are needed to test this hypothesis.

\section*{ACKNOWLEDGMENTS}

We gratefully acknowledge use of the Vizier catalogue (CDS, France),
of data products from the 2MASS survey, and of archival HST and {\em
Chandra} observations obtained from the STScI and SAO websites.

\end{document}